\documentclass[aps,preprint,preprintnumbers,nofootinbib,showpacs]{revtex4}
\usepackage{graphicx,color,amsmath}
\begin{document}
\title{Determination of Nucleon Form factors from Baryonic $B$ Decays}
\author{C. Q. Geng and Y. K. Hsiao}
\affiliation{Department of Physics, National Tsing Hua University,
Hsinchu, Taiwan 300}
\date{\today}
\begin{abstract}
It is the first time that the study of three-body baryonic $B$
decays offers an independent determination of the nucleon form
factors for the time-like four momentum transfer $(t>0)$ region, such as $G^{p,n}_M(t)$ of vector currents
and those of axial ones.
Explicitly, from the data of
$\bar B^0\to n\bar p D^{*+}$ and $\Lambda\bar p\pi^+$ we find
 a constant ratio of $G^n_M(t)/G^p_M(t)=-1.3\pm0.4$,
 which supports
the FENICE experimental result.
The vector and axial-vector form factors of
$p\bar n$, $p\bar p$ and $n\bar n$ pairs due to weak currents are also
presented,
 which can be tested in future experiments.

\end{abstract}
 \maketitle
%=========================================================================
\newpage
%=========================================================================
The nucleon form factors still attract attentions as they reveal
the hadron structures and play important roles in
any scattering or decaying processes
involving baryons \cite{spacelikedata2}.
These form factors depend on
the four momentum transfer $(t)$, which is either
space-like ($t<0$) or time-like ($t>0$).
The behaviors of the form
factors versus $t$ have been extensively studied in various
 QCD models \cite{dispersion,dispersion3,dispersion1,dispersion1-2,Brodsky2,Chernyak,VMD,VMD-2,Tomasi,Tomasi2,DeFalco,more}, such as perturbative QCD (PQCD),
chiral perturbation theory (ChPT),
vector meson dominance (VMD) approach and dispersion
relation (DR) method.
Experimentally,
 accurate electromagnetic data
for vector form factors with the space-like momentum transfer have become abundant \cite{spacelikedata1},
whereas the data on axial-vector form
factors are available only for the space-like region with $|t|<1$ GeV$^2$ in the neutrino-nucleon scattering
\cite{spacelikedata2}. In particular,
the time-like electromagnetic  form factors of the proton have been extracted
from $e^+e^-\to p\bar p$ ($p\bar p\to e^+e^-$) \cite{eetopp},
but only a few data points have been collected for those of the neutron
 by the
FENICE Collaboration \cite{eetonn}.  Currently, due
to experimental difficulties,  there are no data on
 the time-like axial
structures, induced from the weak
currents due to $W$ and $Z$ bosons.
Moreover, there exists some inconsistency
between the measurement and theory
\cite{dispersion,dispersion3,dispersion1,dispersion1-2,Brodsky2,Chernyak,VMD,VMD-2,Tomasi,Tomasi2,DeFalco,more} for the $n\bar n$ data, unlike
the $p\bar p$ case, which seems to be well understood by the theoretical calculations.
Clearly, more theoretical studies  as well as precise experimental measurements
on the nucleon
form factors are  needed to improve our
understanding of strong interactions.

In this paper, we shall show that
the three-body baryonic $B$ decays of $B\to {\bf B}\bar {\bf B'}M$,
such as $\bar B^0\to n\bar p D^{*+}$ and
 $\bar B^0\to\Lambda\bar p\pi^+$,
can provide valuable information on the nucleon form factors.
In general, the three-body baryonic $B$ decays involve time-like form factors
from vector, axial vector, scalar and pseudoscalar currents, respectively. In
the scale of $m_b\sim 4$ GeV, the PQCD
is suitable for us to systematically examine
not only the form
factors of vector currents but also those of axial vector
ones. We note that  the PQCD approach in a
series of works in Refs.
\cite{HouSoni,ChuaHouTsai,ChuaHou,HY,HY2,Geng,HYreview}
has been developed as a reliable tool to explain the experimental data on the baryonic $B$ decays.

In the wildly used factorization method \cite{BW,BSW}, which
splits the four quark operators into two currents by the vacuum
insertion, there are three types of three-body baryonic
$B$ decays: Type I is for the decay in which
a meson is transformed from $B$ together with
an emitted  baryon pair; Type II is for the mode in which
a baryon pair is transited from $B$ together with an ejected meson;
and Type III is the mixture of Types I and II.
With the factorization method, the decay amplitudes for Types I and II are proportional to
 $\langle{\bf B}{\bf \bar
B'}|J^1_\mu|0\rangle\langle M|J_2^\mu|B\rangle $
and
 $\langle M|J^1_\mu|0\rangle\langle{\bf B}{\bf \bar
B'}|J_2^\mu|B\rangle $, respectively.
 For the present measured modes, for instance,
 $\bar B^0\to n\bar p D^{*+}$
\cite{Dstarpn} and $\bar B^0\to\Lambda \bar p \pi^+$
\cite{Lambdappi} belong to Type I, while $\bar B^0\to p\bar p
D^{(*)0}$ \cite{Dpp} and $B^-\to \Lambda \bar p J/\Psi$
\cite{JLambdap} are classified as Type II, whereas $B^-\to p\bar p
K^{(*)-}$, $\bar B^0\to p\bar p K_S$, $B^-\to p\bar p \pi^-$
\cite{pph} and $B^-\to \Lambda\bar \Lambda K^-$
\cite{LambdaLambdaK} are of Type III.
 Although
the decay modes of Types I and III are of our current interest,
those of Type III are inevitably affected by the uncertainties of
the $B\to {\bf B}{\bf \bar B'}$ transition form factors.
In this study, we shall concentrate
on the Type I modes of $\bar B^0\to n\bar p D^{*+}$ and $\bar
B^0\to\Lambda \bar p \pi^+$.

 With the effective Hamiltonians \cite{Hamiltonian} at the
quark level,
% and the assumption of the factorization,
 the decay amplitudes are given by
 \cite{ChuaHouTsai,ChuaHou,HY,HY2,Geng}
\begin{eqnarray}\label{amp}
{\cal A}(\bar B^0\to n\bar p D^{*+})&=&\frac{G_F}{\sqrt
2}V_{cb}V_{ud}^* a_1\langle n\bar p|(\bar d
u)_{V-A}|0\rangle\langle D^{*+}|(\bar c b)_{V-A}|\bar
B^0\rangle\;, \\
{\cal A}(\bar B^0\to\Lambda \bar p \pi^+)&=&\frac{G_F}{\sqrt
2}\bigg\{ (V_{ub}V_{us}^* a_1-V_{tb}V_{ts}^*a_4)\langle \Lambda
\bar p|(\bar s u)_{V-A}|0\rangle
\langle \pi^+|(\bar u b)_{V-A}|\bar B^0\rangle\nonumber\\
%&&-V_{tb}V_{ts}^*a_4\langle \Lambda \bar p|(\bar s u)_{V-A}|0\rangle\langle \pi^+|(\bar u b)_{V-A}|B\rangle\nonumber\\
&& +V_{tb}V_{ts}^*2a_6\langle \Lambda \bar p|(\bar s
u)_{S+P}|0\rangle\langle \pi^+|(\bar u b)_{S-P}|\bar
B^0\rangle\bigg\}\;,
\end{eqnarray}
where  $G_F$ is the Fermi constant, $V_{q_iq_j}$ are the CKM matrix elements, $(q_iq_j)_{V-A}=q_i\gamma_\mu(1-\gamma_5)q_j$,
$(q_iq_j)_{S\pm P }=q_i(1\pm\gamma_5)q_j$ and
$a_i$ ($i=1,4,6$)
are given by
\begin{eqnarray}\label{a146}
a_1=c_1^{eff}+\frac{1}{N_c^{eff}}c_2^{eff}\;,\;a_4=c_4^{eff}+\frac{1}{N_c^{eff}}c_3^{eff}\;,\;a_6=c_6^{eff}+\frac{1}{N^{eff}_c}c_5^{eff}\;,
\end{eqnarray}
with $c_i^{eff}\;(i=1,2,\cdots, 6)$ being the effective Wilson
coefficients (WCs) shown in Refs. \cite{Hamiltonian} and $N_c^{eff}$
the  effective color number.
%$a_i=c^{eff}_i+c^{eff}_{i\pm 1}/N_c^{eff}$ for $i=$odd (even)
%in terms of the effective Wilson coefficients $c^{eff}_i$ and the
%effective color number $N_c^{eff}$. 
Here, we have used the generalized 
factorization method with the non-factorizable effect absorbed in
$N_{c}^{eff}$.
For the matrix elements of
 $\langle \pi^+|(\bar u
b)_{V-A}|\bar B^0\rangle$ and $\langle D^{*+}|(\bar c
b)_{V-A}|\bar B^0\rangle$,
we use the results
in Refs. \cite{BSW,Dstarff}. For the
time-like baryonic form factors, we have
\begin{eqnarray}\label{form3} \langle {\bf
B}{\bf\bar B'}|\bar q_i\gamma_\mu q_j|0\rangle &=&\bar u(p_{\bf
B})\bigg\{F_1(t)\gamma_\mu+\frac{F_2(t)}{m_{\bf B}+m_{\bf \bar
B'}}
i\sigma_{\mu\nu}(p_{\bf \bar B'}+p_{\bf B})_\mu\bigg\}v(p_{\bf \bar B'})%\;,
\nonumber\\
&=& \bar u(p_{\bf B})\bigg\{[F_1(t)+F_2(t)]\gamma_\mu+\frac{F_2(t)}{m_{\bf B}+m_{\bf \bar B}}
(p_{\bf \bar B}-p_{\bf B})_\mu\bigg\}v(p_{\bf \bar B})\;,\nonumber\\
\langle {\bf B}{\bf\bar B'}|\bar q_i\gamma_\mu\gamma_5
q_j|0\rangle&=&\bar u(p_{\bf
B})\bigg\{g_A(t)\gamma_\mu+\frac{h_A(t)}{m_{\bf B}+m_{\bf \bar
B'}} (p_{\bf \bar B'}+p_{\bf B})_\mu\bigg\}\gamma_5 v(p_{\bf \bar
B})\,,
\nonumber\\
\langle {\bf B}{\bf\bar B'}|\bar q_i q_j|0\rangle &=&f_S(t)\bar
u(p_{\bf B})v(p_{\bf\bar B'})\;,
\nonumber\\
 \langle {\bf B}{\bf\bar
B'}|\bar q_i\gamma_5 q_j|0\rangle &=&g_P (t)\bar u(p_{\bf
B})\gamma_5 v(p_{\bf\bar B'})\,,
\end{eqnarray}
where the four momentum transfer in the time-like region is
$t=(p_{\bf B}+p_{\bf\bar B'})^2$, $q_i=u$, $d$ and $s$, and
$F_1$, $F_2$, $g_A$, $h_A$, $f_S$ and $g_P$ are the form factors.

In this paper, we will study the form factors
in Eq. (\ref{form3})
based on the experimental data in the baryonic B decays.
We begin by defining the baryonic form factors of $F_1$ and $g_A$
by
\begin{eqnarray}\label{F1gA}
\langle {\bf B}|J_\mu^{em}|{\bf B'}\rangle=\bar u(p_{\bf
B})\bigg[F_1(t)\gamma_\mu+g_A(t)\gamma_\mu\gamma_5\bigg]v(p_{\bf B'})\;,
\end{eqnarray}
where $J_\mu^{em}=Q_q\bar{q}\gamma_\mu\,q$ and
$t=(p_{\bf B}-p_{\bf B'})^2$. Note that
$F_2$ and $h_A$ are not included in Eq. (\ref{F1gA})
due to the helicity
conservation.
To exhibit the chirality or helicity, we rewrite Eq.
(\ref{F1gA})  as
\begin{eqnarray}\label{Gff1}
\langle {\bf B}_{\uparrow+\downarrow}|J_\mu^{em}|{\bf
B'}_{\uparrow+\downarrow}\rangle=\bar u(p_{\bf B})\bigg[\gamma_\mu
\frac{1+\gamma_5}{2}G^\uparrow(t)+\gamma_\mu
\frac{1-\gamma_5}{2}G^\downarrow(t)\bigg]u(p_{\bf B'})\;,
\end{eqnarray}
where $|{\bf B}_{\uparrow+\downarrow}\rangle\equiv |{\bf
B}_{\uparrow}\rangle+|{\bf B}_{\downarrow}\rangle$ respects both flavor
$SU(3)$  and  spin $SU(2)$ symmetries,
{\it e.g.}, $|p_\uparrow \rangle=\sqrt{1/18}[u_\uparrow
u_\downarrow d_\uparrow+u_\downarrow u_\uparrow
d_\uparrow-2u_\uparrow u_\uparrow d_\downarrow+permutations]$.
In Eq. (\ref{Gff1}),
$G^\uparrow(t)$ and $G^\downarrow(t)$ represent the  right-handed
and left-handed form factors,  which can be
further decomposed as
\begin{eqnarray}\label{Gff2}
G^\uparrow(t)=e^\uparrow_{||}G_{||}(t)+e^\uparrow_{\overline{||}}G_{\overline{||}}(t)\;,\;\;
G^\downarrow(t)=e^\downarrow_{||}G_{||}(t)+e^\downarrow_{\overline{||}}G_{\overline{||}}(t)\;,
\end{eqnarray}
where the constants $e^{\uparrow(\downarrow)}_{||}$ and
$e^{\uparrow(\downarrow)}_{\overline{||}}$ are defined by
\begin{eqnarray}\label{Gff3}
e^{\uparrow(\downarrow)}_{||}=\langle {\bf
B_{\uparrow(\downarrow)}}|{\bf Q_{||}}|{\bf
B'_{\uparrow(\downarrow)}}\rangle\;,\;\;
e^{\uparrow(\downarrow)}_{\overline{||}}=\langle {\bf
B_{\uparrow(\downarrow)}}|{\bf Q_{\overline{||}}}|{\bf
B'_{\uparrow(\downarrow)}}\rangle\;,
\end{eqnarray}
respectively,
with ${\bf Q_{||(\overline{||})}}=\sum_i
Q_{||(\overline{||})}(i)$.
%where $i=1,2,3$ are the indices of the valence quarks.
In Eq. (\ref{Gff3}), the summation is
%they are summing 
over the charges carried by
the valence quarks $(i=1,2,3)$ in
the baryon with helicities parallel $(||)$ and anti-parallel
$(\overline{||})$ to the baryon spin directions of
$(\uparrow,\downarrow)$. Since $G_{||(\overline{||})}(t)$
are the form factors
accompanied by the
(anti-)parallel hard-scattering amplitudes with baryon wave
functions,
 based on the QCD counting rules in
the PQCD \cite{Brodsky1,Brodsky2,Brodsky3}, they can be expressed by
\begin{eqnarray}\label{Gff4}
G_{||}(t)=\frac{C_{||}}{t^2}\left[\text{ln}(\frac{t}{\Lambda^2_0})\right]^{-\gamma}\;,\;\;
G_{\overline{||}}(t)=\frac{C_{\overline{||}}}{t^2}\left[\text{ln}(\frac{t}{\Lambda^2_0})\right]^{-\gamma}\;,
\end{eqnarray}
where $\gamma=2.148$, $\Lambda_0=300$ MeV \cite{dispersion1} and
$C_{||,\overline{||}}$ are parameters to be determined.
%It is noted that $e^{\uparrow(\downarrow)}_{||,\overline{||}}$
% represent flavor and spin symmetries.
We remark that the asymptotic formulas in Eq. (\ref{Gff4}) are
exact only when $t$ is large. For smaller $t$, such as when $t$ being
close to the two-nucleon threshold, higher power corrections are expected \cite{GH-high}. The corrections will be averaged into the
errors of $C_{||,\overline{||}}$ in our data fitting.
 From Eqs. (\ref{F1gA})-(\ref{Gff3}),
   we have
\begin{eqnarray}
\label{form9}
F_1(t)&=&(e^\uparrow_{||}+e^\downarrow_{||})G_{||}(t)+(e^\uparrow_{\overline{||}}+e^\downarrow_{\overline{||}})G_{\overline{||}}(t)\;,\nonumber\\
g_A(t)&=&(e^\uparrow_{||}-e^\downarrow_{||})G_{||}(t)+(e^\uparrow_{\overline{||}}-e^\downarrow_{\overline{||}})G_{\overline{||}}(t)\;.
\end{eqnarray}
Although the above equations are derived in the space-like region,
the time-like form factors can be easily written via the crossing
symmetry \cite{dispersion1,Tomasi3}, which transforms the particle in the initial
state to its anti-particle in the final state and reverses its
helicity. However, in general, the values of the
time-like $G_{||}(t)$ and
$G_{\overline{||}}(t)$ form factors are complex numbers
unlike the space-like ones for which there is a time reversal symmetry between initial and final states.

In Eq. (\ref{form3}),
$F_2$ is suppressed by
$1/(t\text{ln}[t/\Lambda_0^2])$ in comparison with $F_1$ \cite{F2,F2b} and therefore
can be safely ignored,
while $g_P$ is found to be related to $f_S$ as \cite{ChuaHou}
\begin{eqnarray}\label{gp}
g_P=f_S\;,
\end{eqnarray}
 and $f_S$ and $h_A$ are deduced from equation of
motion, given by
\begin{eqnarray}\label{fsha}
f_S(t)&=&\frac{m_{\bf B}-m_{\bf B'}}{m_{q_i}-m_{q_j}}F_1(t)\;,
\nonumber\\
h_A(t)&=&-\frac{(m_{\bf B}+m_{\bf B'})^2}{t}g_A(t)\;.
\end{eqnarray}
Thus, once we figure out $F_1$ and $g_A$,
all other form factors
$f_S$, $h_A$ and $g_P$ will be determined
 in terms of Eqs.
(\ref{gp}) and (\ref{fsha}).
For the electromagnetic current $J^{em}_\mu=\frac{2}{3}\bar
u\gamma_\mu u-\frac{1}{3}\bar d\gamma_\mu d$,
from Eq. (\ref{form9}) it is clear that $g_A=0$ since
$e^\uparrow_{||(\overline{||})}=e^\downarrow_{||(\overline{||})}$.
Furthermore, from Eqs. (\ref{form3})-(\ref{Gff3}) we have
\begin{eqnarray}\label{em}
G^p_M(t) &\equiv& F_1^{p\bar{p}}(t)+F_2^{p\bar{p}}(t)
\;\simeq\; F^{p\bar
p}_{1(em)}(t)\;=\;G_{||}(t)\;,
\nonumber\\
G^n_M(t)
&\equiv& F_1^{n\bar{n}}(t)+F_2^{n\bar{n}}(t)
\;\simeq\; F^{n\bar
n}_{1(em)}(t)\;=\;-\frac{G_{||}(t)}{3}+\frac{G_{\overline{||}}(t)}{3}\,,
\end{eqnarray}
where we have neglected the small $F_2^{N\bar{N}}$ terms
comparing with those of $F_1^{N\bar{N}}$ \cite{F2,F2b}.
Similarly, for the $Z$ coupled current  $J^Z_\mu=\frac{1}{2}\bar
u\gamma_\mu(\frac{1-\gamma_5}{2}) u-\frac{1}{2}\bar
d\gamma_\mu(\frac{1-\gamma_5}{2}) d-\sin^2\theta_W J^{em}_\mu$,
we use the same  form factors defined in Eqs. (\ref{F1gA})
and (\ref{Gff1}) by replacing $J_\mu^{em}$ with $J_\mu^{Z}$ and
we get
\begin{eqnarray}\label{nucleon}
F^{p\bar p}_{1(Z)}(t)&=&
\frac{2-3\sin^2\theta_W}{3}G_{||}(t)-\frac{1}{6}G_{\overline{||}}(t)\;,                                    \ \ \ \ \ \ g_{A(Z)}^{p\bar p}(t)=\frac{2}{3}G_{||}(t)+\frac{1}{6}G_{\overline{||}}(t)\;,
\nonumber\\
 F^{n\bar n}_{1(Z)}(t)&=&\frac{-2-\sin^2\theta_W}{3}G_{||}(t)+\frac{1+2\sin^2\theta_W}{6}G_{\overline{||}}(t),\;
 g_{A(Z)}^{n\bar n}(t)=-\frac{2}{3}G_{||}(t)-\frac{1}{6}G_{\overline{||}}(t)\,.
\end{eqnarray}
 Since the behaviors
of $G_{||}(t)$ and $G_{\overline{||}}(t)$ have been given in Eq.
(\ref{Gff4}), what we shall do next is to fix the parameters
$C_{||}$ and $C_{\overline{||}}$
 in terms of
 \begin{eqnarray}\label{weak}
%\begin{array}{rr}
F^{n\bar p}_{1}(t)&=&\frac{4}{3}G_{||}(t)-\frac{1}{3}G_{\overline{||}}(t)\;,\ g_{A}^{n\bar p}(t)=\frac{4}{3}G_{||}(t)+\frac{1}{3}G_{\overline{||}}(t)\;,
\nonumber\\
                 F^{\Lambda\bar p}_{1}(t)&=&\sqrt{\frac{3}{2}}G_{||}(t)\;,\;\;\;\;\;\;\;\;\;\;\;\ \ g_{A}^{\Lambda\bar p}(t)=\sqrt{\frac{3}{2}}G_{||}(t)\;,
%\end{array}\bigg\}
\end{eqnarray}
derived from Eqs. (\ref{amp})-(\ref{Gff3}) with the data in
$\bar B^0\to n\bar p D^{*+}$ and
 $\bar B^0\to\Lambda\bar p\pi^+$.
 
 Before performing the numerical analysis, we would like briefly
 discuss the generalized factorization method.
 %Even though the baryonic form factors via different currents can
%be studied in three-body baryonic $B$ decays, the based
It is known that the
factorization method \cite{BW,BSW}
%, which splits the four quark
%operators into two currents by the vacuum insertion in analogy
%with semileptonic meson decays,
suffers from several possible
hadron uncertainties, such as 
%The uncertainties are 
those from the nonfactorizable
effect, annihilation contribution and final state interaction.
To describe these uncertainties, we take the decay of $\bar B^0\to n\bar p D^{*+}$ as an example, while those for
 $\bar B^0\to \Lambda\bar p\pi^+$ can be treated in a similar manner.
The amplitude of $\bar B^0\to n\bar p D^{*+}$ from the color
suppressed operator is given by
\begin{eqnarray}
&&c_2^{eff}\langle n\bar p D^{*+}|(\bar d_\alpha u_\beta)_{V-A}(\bar c_\beta b_\alpha)_{V-A}|\bar B^0\rangle\nonumber\\
&&=\frac{c_2^{eff}}{N_c}\langle n\bar p|(\bar d u)_{V-A}|0\rangle\langle
D^{*+}|(\bar c b)_{V-A}|\bar B^0\rangle +\frac{c_2^{eff}}{2}\langle n\bar p
D^{*+}|(\bar d \lambda^a u)_{V-A}(\bar c \lambda^a b)_{V-A}|\bar B^0\rangle\,,
\end{eqnarray}
where $\delta_{\beta\beta'}\delta_{\alpha\alpha'}=
\delta_{\beta\alpha}\delta_{\alpha'\beta'}/N_c+\lambda^a_{\beta\alpha}\lambda^a_{\alpha'\beta'}/2$
has been used to deal with color index $\alpha$ ($\beta$)
 and the
second term on the right-hand side is the so-called
nonfactorizable effect. Although the nonfactorizable effect cannot
be directly and unambiguously determined by theoretical
calculations, in the generalized factorization method
\cite{Hamiltonian}, this contribution can be 
absorbed in
%parameterizing $N_c$  as 
the effective color
number $N_c^{eff}$ running from 2 to $\infty$
in Eq. (\ref{a146}). 
%It is acceptable
%since the nonfactorizable effect should be regarded as a QCD
%effect, maximally to the order of its main corrected target.\\
The amplitude of the annihilation contribution is given by
\begin{eqnarray}\label{ampanni}
{\cal A}_{an}(\bar B^0\to n\bar p D^{*+})&=&\frac{G_F}{\sqrt
2}V_{cb}V_{ud}^* a_2\langle n\bar p D^{*+}|(\bar c
u)_{V-A}|0\rangle\langle 0|(\bar d b)_{V-A}|\bar B^0\rangle\;,
\end{eqnarray}
where $a_2$ is around $(0.1-0.01)a_1$, which is suppressed.
In addition,
%Besides, since 
based on the power expansion of $1/t$ in the PQCD approach,
%which accounts for two gluon propagators to form baryon pair
%\cite{ChuaHouTsai2, AngdisppK, Brodsky1, Brodsky2, Brodsky3},
%successfully explains the threshold enhancement from the data of
%spectra in three-body baryonic $B$ decays, in the same sense,
${\cal A}_{an}(\bar B^0\to n\bar p D^{*+})\propto 1/t^3$, which 
is much suppressed than ${\cal A}(\bar B^0\to n\bar p D^{*+})$ 
as
%due to the fact that the squared transmitted energy 
%$t$ is in the order of 
$t\sim m^{2}_{b}$ \cite{CPppK}.
For the final state interaction, the most possible source is via
the two-particle rescattering to the baryon pair, such as $\bar B^0\to
M_1 M_2 D^{*+}\to n\bar p D^{*+}$ with $M_{1,2}$ representing
meson states. However, such processes would shape the curve
associated with the phase spaces in the decay rate distributions
\cite{LambdaCBELLE,LambdaCChen,LambdaCCheng}, which have been
excluded in the charmless baryonic $B$ decay experiments.
 
%Although the nonfactorizable effects are in a limited
%order, while the contributions from the 
%annihilation and final state interaction should be small,
%to really understand the three-body baryonic $B$ decays,
%the uncertainties should be at least numerically under control.
%

In our numerical analysis, we take 
$c_i^{eff}(i=1, 2,\cdots, 6)\simeq
(1.17,-0.37,0.0246,-0.0523,0.0154,-0.066)$ 
\cite{Hamiltonian,Ali,Buras},
%$(a_1,a_4,a_6)\simeq (1.05,-0.0441,-0.0609)$ 
$m_u(m_b)=3.2$ MeV, $m_s(m_b)=90$ MeV
\cite{Hamiltonian} and $m_b(m_b)=4.2$ GeV \cite{mbmass},
and the weak phase $\gamma=59.8^\circ\pm 4.9^\circ$ \cite{pdg}.
%while the Fermi constant and the CKM matrix elements with errors
%can be found in Ref. \cite{pdg}.
%%We note that in the generalized factorization approach
%%$a_{i}$ are physical parameters and scale independent 
%%\cite{Hamiltonian,Ali}.
For the
%With the 
experimental data in the B decays, we use
\cite{Dstarpn,Lambdappi} 
\begin{eqnarray}
Br(\bar B^0\to n\bar p D^{*+})&=&(14.5\pm 4.3)\times 10^{-4}\,,
\nonumber\\
%and
Br(\bar B^0\to \Lambda\bar p\pi^+)&=&(3.29\pm 0.47)\times
10^{-6}\,,
\label{ExptBRs}
\end{eqnarray}
and  the other available data in Ref. \cite{Lambdappi}, such as
 spectrum vs. invariant mass and
angular distributions.

Based on the $\chi^2$ fitting,
%which statistically reflects the uncertain ranges.
we obtain
\begin{eqnarray}
\chi^2/dof=0.7\,,\;1/N_c^{eff}=0.2\pm 0.2,
\end{eqnarray}
where dof denotes the degree of freedom. 
It is clear that $\chi^2/dof=0.7$ presents a
reliable fit, while $N_c^{eff}\sim 2.5-\infty$ means a limited
nonfactorizable effects with the small contributions from the annihilation and final state interaction. We stress that if the hadronic
uncertainties were large, $N_c^{eff}$ from 2 to $\infty$ would
not be account for the data.
The coefficients of $C_{||}$ and $C_{\overline{||}}$
in Eq. (\ref{Gff4}) are found to be
\begin{eqnarray}\label{C1}
C_{||}=83.7\pm 5.7\,GeV^{4}\;\;\text{and}\;\;C_{\overline{||}}=-246.3\pm 92.1\,GeV^{4}
\end{eqnarray}
which lead to
\begin{eqnarray}\label{Ratio}
G_{\overline{||}}(t)/G_{||}(t)&=&-2.9\pm 1.1\,.
\end{eqnarray}
%Note that the ratio in Eq. (\ref{Ratio}) is a constant as
Here, we have assumed that both $C_{||}$ and $C_{\overline{||}}$
are real since
their imaginary parts
are expected to be
 small based on the argument in Refs.
\cite{Tomasi2,Tomasi3} as well as the result in the DR method \cite{dispersion2}.

\begin{figure}[t!]
\centering
\includegraphics[width=1.89in]{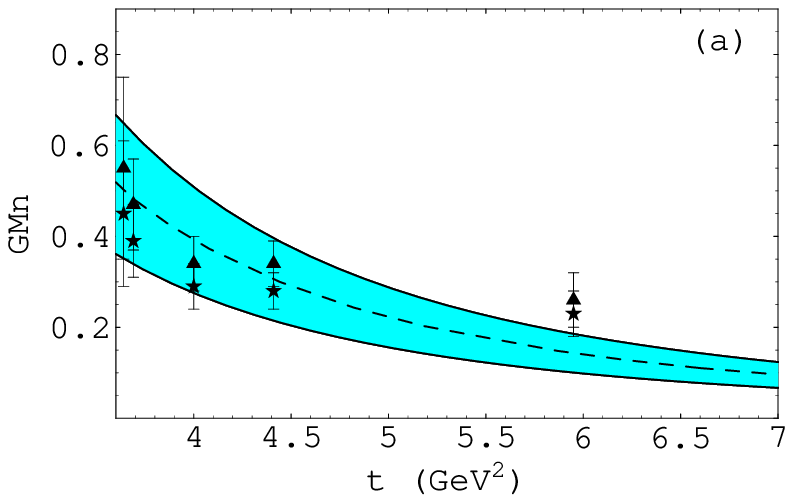}
\includegraphics[width=2.0in]{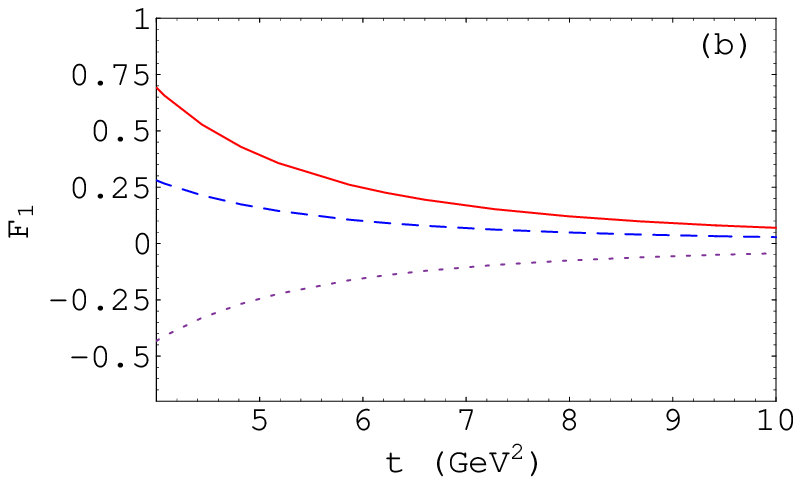}
\includegraphics[width=2.0in]{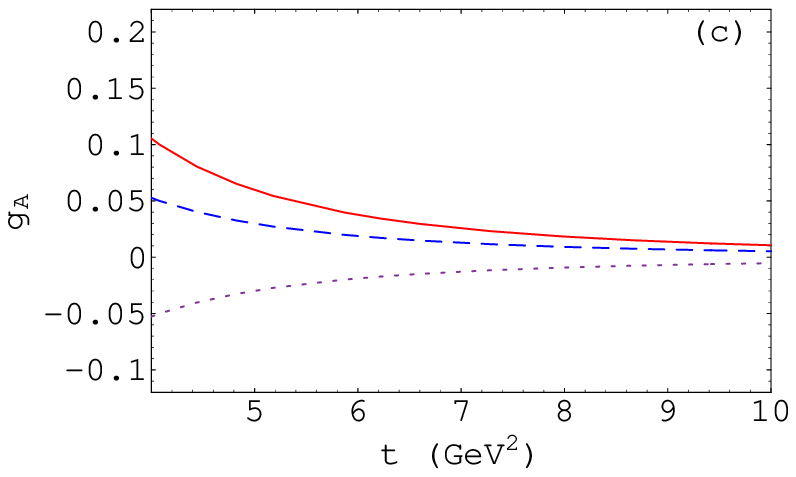}
\caption{\label{fig}
 Form factors of (a) $G^n_M(t)$,
  (b) $F_1(t)$ and (c) $g_A(t)$ with
  the time-like four momentum transfer $t$,
  where  the star and triangle symbols represent the FENICE
data \cite{eetonn} with the assumptions of  $|G_M^n|=|G_E^n|$ and
$|G_E^n|=0$, and  the solid, dash
and dotted curve stand for (b) $F^{p\bar n}_1(t)$, $F^{p\bar
p}_{1(Z)}(t)$ and $F^{n\bar n}_{1(Z)}(t)$ and (c) $g^{p\bar
n}_A(t)$, $g^{p\bar p}_{A(Z)}(t)$ and $g^{n\bar n}_{A(Z)}(t)$,
respectively.}
\end{figure}

As seen in Fig. \ref{fig}a,
the fitted values of $G^n_M(t)$ extracted from the
$B$ decays are in agreement with   the
FENICE data \cite{eetonn} by the assumptions of  $|G_M^n|=|G_E^n|$ and $|G_E^n|=0$.
We note that $G^n_E(t)\equiv F_1^{n\bar{n}}(t)+{t\over 4M_n^2}F_2^{n\bar{n}}$ is the neutron electric form factor. Clearly, in our calculation based on the QCD counting rule, $G_E^n\sim G_M^n$.

 From Eqs. (\ref{Gff4}),  (\ref{em}) and (\ref{C1}), we get
\begin{eqnarray}
%\label{Result} G^n_M(t)/G^p_M(t)&=&-1.1_{-0.3}^{+0.4}\,,
\label{Result} G^n_M(t)/G^p_M(t)&=&-1.3\pm 0.4\,,
\end{eqnarray}
which supports the measurements in
Refs. \cite{eetopp} and \cite{eetonn}. We note that 
 in Eq. (\ref{Result}) the ratio is a constant due to
the same power expansions in Eq. (\ref{Gff4}) and
the minus sign 
%in Eq. (\ref{Result}) 
is necessary in order to match the measured  baryonic $B$
decay branching ratios, which also
confirms the theoretical result in
 Refs. \cite{Brodsky1,Brodsky2,Brodsky3,Chernyak}.
Moreover, the
puzzle
%since 1998 for the ratio
of $|G^n_M(t)/G^p_M(t)|\geq 1$
is also solved as indicated in Eq. (\ref{Result}).
We note that the ratio
$G^n_M(t)/G^p_M(t)$ is predicted to be $-2/3$ and $-1/2$ in the QCD
counting~\cite{Brodsky2} and  sum rules~\cite{Chernyak}, respectively,
while the DR
 \cite{dispersion,dispersion3,dispersion1,dispersion1-2} and VMD
 \cite{VMD,VMD-2} methods yield only
 half of the values indicated by the data points.
It is interesting to see that
  if $G_{||}(t)/G_{\overline{||}}(t)= -1$ instead of the fitted one in
  Eq. (\ref{Ratio}),
from
Eq. (\ref{em})
 $G^M_P(t)/G^p_M(t)=-2/3$ is recovered as in
Ref.~\cite{Brodsky2}.
Therefore, our result on the ratio
 in Eq. (\ref{Ratio})
could help us
to improve the PQCD calculations.

Since we have related all the nucleon
form factors as in Eqs. (\ref{em}), (\ref{nucleon}) and
(\ref{weak}), we find
\begin{eqnarray}
\label{F1gAR}
F^{n\bar p}_1(t)&=&(193.7\pm 31.6)G_{p}(t)\;,\;\;\;\;\;\;g^{n\bar p}_A(t)=(29.5\pm 31.6)G_{p}(t)\;,\;\nonumber\\
F^{p\bar p}_{1(Z)}(t)&=&(78.4\pm 15.6)G_{p}(t)\;,\;\;\;\;\;g^{p\bar p}_{A(Z)}(t)=(14.8\pm 15.8)G_{p}(t)\;,\;\nonumber\\
F^{n\bar n}_{1(Z)}(t)&=&(-121.1\pm 22.5)G_{p}(t)\;,\;g^{n\bar n}_{A(Z)}(t)=(-14.8\pm 15.8)G_{p}(t)\;,\;
\end{eqnarray}
where $G_{p}(t)\equiv 1/{t^2}[\text{ln}(t/\Lambda^2_0)]^{-\gamma}$.
The central values of $F_1^{N\bar{N'}}$ and $g_A^{N\bar{N'}}$
in Eq. (\ref{F1gAR}) are shown
in Figs. \ref{fig}b and \ref{fig}c, respectively.
The valid
ranges for these time-like form factors
are the same as those in the three-body baryonic $B$ decays of
$B\to {\bf B\bar{B}'}M$,
$i.e.$,
$(m_{\bf B}+m_{\bf B'})^2\simeq 4\;\text{GeV}^2\leq t\leq
(m_B-m_M)^2\simeq 16-25\;\text{GeV}^2$.

In sum, we have shown that the study of
the measured three-body baryonic $B$
decays of $\bar B^0\to n\bar p D^{*+}$ and $\bar B^0\to
\Lambda\bar p\pi^+$ leads to
 $G^n_P(t)/G^p_t(t)=-1.3\pm0.4$,
which supports the FENICE measurement.
 The minus sign for the ratio
 $G^n_P(t)/G^p_t(t)$,
given  by the previous theoretical calculations, has been enforced to fit
the $B$ decay data.
We have  pointed out that our fitted value for the
ratio of
$G_{||}(t)$ and $G_{\overline{||}}(t)$
may be useful for us to perform various QCD calculations on
$e^+e^-\to n\bar n$.
We have also predicted the time-like vector and
axial vector nucleon  form factors induced from the weak currents,
such as $F^{N\bar N'}_1(t)$
 and $g^{N\bar N'}_A(t)$
 ($N,N'=p$ and $n$).
 
Finally, we remark that
apart from the use of $\bar B^0\to p\bar n D^{*+}$ and $\bar
B^0\to \Lambda\bar p\pi^+$, there are more decays directly
connecting to the time-like form factors, such as $\bar B^{0}\to
p\bar n (D^{+},\,\rho^+,\,\pi^+)$ and $\bar B^0\to \Lambda\bar p\rho^+$
as well as the corresponding charged $B$ modes, which are within the accessibility of the current $B$ factories at KEK and SLAC.
It is clear that as more and more data available from
current and future $B$ factories,
the nucleon form factors can be further constrained and determined. Moreover, the new measurements in $e^+ e^-\to n\bar n$
are  progressing in DA$\Phi$NE at Frascati \cite{VMD} and
planning  in
PANDA and PAX  at GSI \cite{dispersion1-2}.
 As for the weak nucleon form
factors, since the scattering of $e^+e^-$ at BABAR
is at the $m_B$ scale, $F^{N\bar N'}_1(t)$ and
$g^{N\bar N'}_A(t)$ ($N,N'=p$ and $n$) can be studied via the
left-right helicity asymmetry \cite{GH-high} of
$A_{PV}=(d\sigma_R-d\sigma_L)/(d\sigma_R+d\sigma_L)$ as in the SAMPLE
experiment~\cite{spacelikedata2}.\\

%\section*{Acknowledgements}
%We would like to thank Prof. ** for useful discussions.
 This work
is financially supported by the National Science Council of
Republic of China under the contract
\#s NSC-94-2112-M-007-(004,005).

%\newpage

\end{document}